\documentclass{hep99}
\usepackage{psfig}

\newlength{\figwidth}
\newcommand{\eabe} {\begin{eqnarray}}
\newcommand{\eaen} {\end{eqnarray}}
\newcommand{\eqbe} {\begin{equation}}
\newcommand{\eqen} {\end{equation}}

\newcommand{\mrm} {\mathrm}
\newcommand{\srm}[1] {_{\mathrm{#1}}}
\newcommand{\ol} {\overline}

\newcommand{\ra} {\rightarrow}

\newcommand{\bibl}[5]
	{#1, {\it #2} {\bf #3} (#4) #5}

\newcommand{\pair}[1] {${\mrm {#1 \ol #1} }$}

\begin{document}
\title{
\vbox{
	\vskip-60pt\hbox to\hsize{\normalsize
{\mdseries \mbox{NORDITA-99/74 HE}}}
\vskip 35pt}  
Observables Probing the Perturbative-Nonperturbative Transition Region in QCD}

\author{P Ed\'en}

\address{Nordita, Blegdamsvej 17, DK-2100 Copenhagen, Denmark\\
E-mail: {\tt eden@nordita.dk}}

\abstract{
Benefiting from the high statistics from $e^+e^-$ experiments at the ${\mrm Z}^0$ resonance, it is possible to impose strong two-jet cuts on the data without losing the statistical significance. In these events perturbative activity is suppressed and hadronization effects can be more prominent. I give two examples of observables that can be important tools for a more detailed study of the hadronization mechanism.
}

\maketitle

\section{Introduction}\label{sec:intro}
High energy reactions like $e^+e^-$$\ra$ hadrons are usually described in terms of two phases, a perturbative parton cascade followed by soft hadronization described by a phenomenological model. 
A detailed description of the final hadronic state, including e.g.\
flavour, the baryon to meson ratio, spin and polarization as well as
correlations, depends strongly on the
non-perturbative properties of QCD. 
To gain insight in the soft phase of the reaction it is essential to isolate the hadronization process
from the perturbative cascade. 

With the very high statistics available from the LEP1 experiments, it is possible to impose event cuts which exclude a significant amount of perturbative activity, and still have remaining events numerous enough for detailed studies of hadronization.
This is investigated in some detail in~\cite{2jet}, and this talk is a summary of that reference.

The most studied hadronization models are based on string dynamics~\cite{lundstring} or
cluster fragmentation~\cite{herwig}. 
In events with unusually low gluon radiation, the cluster approach must be modified, as some clusters can get very large masses. 
Consequently, if we select events with low gluon activity we
cut away those events for which the cluster model is meant to work best. For this
reason I will focus on different versions of string hadronization.

In this talk I first discuss a set of different event cuts (section~\ref{sec:cuts}) and then  give
two examples of observables that may distinguish different model assumptions (section~\ref{sec:Qt} and~\ref{sec:screw}).

\section{Cuts}\label{sec:cuts}
We first discuss event shape cuts suitable to extract events with little perturbative activity. The performance of the cuts can be investigated using Monte Carlo simulations, where the underlying parton state is known.
We examine the cuts in terms of purity, defined as the rate of events where the highest $k_\perp$ for a gluon emission, $k_{\perp\mrm{max}}$, is below some $k_{\perp0}$,  and efficiency,
 defined as the acceptance rate among the desired events with $k_{\perp\mrm{max}}$$<$$k_{\perp0}$. 

The cuts are applied to hadronic ${\mrm Z}^0$ events generated by the \textsc{Ariadne} and \textsc{Jetset} MC~\cite{ariadne,jetset}. \textsc{Ariadne} is an implementation of the colour dipole formalism for a QCD cascade~\cite{cdm}. This cascade is 
ordered in  $k_\perp$, which implies that $k_{\perp\mrm{max}}$ is easily extracted as the $k_\perp$ of the first emission. \textsc{Jetset} is a MC for the Lund string fragmentation model~\cite{lundstring}.

\begin{figure}[t]
  \begin{center}  \hbox{ \vbox{
	\mbox{\psfig{figure=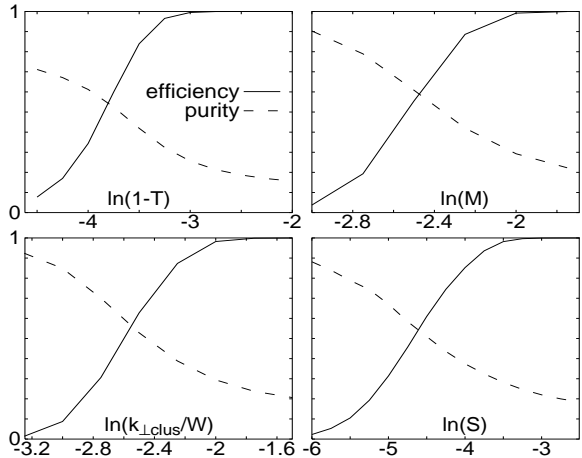,width=0.99\linewidth}}
   }  }
  \end{center}  \caption{\em Purity and efficiency for different two-jet cuts.}
  \label{f:cuts}
\end{figure} 
Fig~\ref{f:cuts} shows efficiency and purity results for $k_{\perp0}$$=2$GeV, for four different two-jet cuts: thrust $T$, thrust major $M$, sphericity $S$ and a jet clustering resolution scale (using the  the \textsc{Diclus} algorithm~\cite{diclus}). 
As seen, the purity and efficiency measures are not very sensitive to the choice of cut. There is thus no optimal cut in all respects, and as we will see, the preferred cut instead depends on the observable under investigation.

\section{Example 1: ${\mathbf p}_\perp$ transfer}~\label{sec:Qt}
One interesting collective variable 
(frequently advocated by E. de Wolf~\cite{WolfQt})
is the vectorial $p_{\perp}$ sum over all hadrons with rapidity less
than a given rapidity $y$
\eqbe{\mathbf Q}_\perp(y)\equiv \sum {\mathbf p}_{\perp i}\Theta(y-y_i), \label{e:Qtdef} \eqen
which measures the $p_\perp$ transfer over the rapidity $y$.

An essential point when using the $Q_\perp$ variable is the choice of axis. 
If the thrust axis is used,
$Q_\perp(y)$ is kinematically constrained to be equal to zero for $y=0$.
This axis is therefore very unsuitable and MC simulations show that the sphericity axis is more appropriate~\cite{2jet}.

The $p_\perp$ transfer observable is sensitive to the  $p_\perp$ correlation length. If $p_\perp$ is locally conserved, average $Q_\perp$ is similar to average $p_\perp$. On the other hand, if only global constraints restrict $p_\perp$, the collective measure $Q_\perp$ gets significantly larger than $p_\perp$. 

Different assumptions about local $p_\perp$ conservation are possible and available on the MC market.
In the \textsc{Jetset} default version of the Lund string fragmentation model the transverse momentum of one hadron is fully compensated by two ``neighbouring'' hadrons which in general are very nearby in rapidity.

Motivated by two-particle correlations observed in hadron--hadron collisions~\cite{K2} a model where only a fraction $\gamma<1$ of the hadron $p_\perp$ is compensated by the neighbouring hadrons is presented in~\cite{jim}.  $\gamma$ is assumed to depend on the hadron mass $M$, and is parametrized as $\gamma=(1+M_0/M)^{-1}$, where $M_0=0.9$GeV. In this ``partial $p_\perp$ compensation'' model the $p_\perp$ correlation length is finite, but larger than in \textsc{Jetset} default.
(A partial $p_\perp$ compensation is also assumed in the UCLA model~\cite{ucla}, with $\gamma=1/2$.)

To test whether the observable $Q_\perp$ can probe a local $p_\perp$ conservation, we also want to study a situation where global $p_\perp$ conservation is the only constraint on hadronic transverse momenta. This can be achieved by 
setting the parameters of the partial $p_{\perp}$ compensation model to get $\gamma \ll 1$ but still keeping the inclusive $p_\perp$ width finite.

In two-jet events, the hadrons can also acquire $p_\perp$ from soft gluons. The compensation of this $p_\perp$ is in general not identical to the one assumed in the fragmentation, and depends on the recoil treatment of the cascade formalism. Thus the assumed $p_\perp$ conservation length may depend on the treatment of recoils in the cascade, and on the cutoff scale, which determines to what extent hadron $p_\perp$ originates from the cascade or the hadronization.

\begin{table}[tb]
  \begin{center} 
    \begin{tabular}{|l||l|l|l|l||}
\hline
 & \multicolumn{2}{c|}{default} & partial & uncorr. \\
\multicolumn{1}{|c||}{$\left<Q_\perp^2\right>$(GeV$^2$)} & \multicolumn{2}{c|}{ } & $p_\perp$ & $p_\perp$\\
& \multicolumn{2}{c|}{ } & comp. &\\
\hline
\multicolumn{1}{|c||}{$k_{\perp\mrm{cut}}$ (GeV)} & 1.5 & 0.6 & 0.6 & 0.6\\
\hline
No cascade & - & 0.60 & 0.71 & 0.74 \\
With cascade: & & & &\\
\multicolumn{1}{|r||}{all events} & 12.2 & 12.3 & 12.4 & 12.6 \\
\multicolumn{1}{|r||}{two-jet events} & 0.86 & 0.94 & 1.02 & 1.09\\
\hline
    \end{tabular}
  \caption{\em Average $Q_\perp^2$, measured w.r.t.\ the sphericity axis in a central rapidity range $|y|<2$. Results for different assumptions about hadron $p_\perp$ correlation lengths. The differences in the predictions are shadowed by the perturbative cascade, but can be restored by a two-jet cut. Here a cut $\ln(S)<-4.5$ has been used.}
  \label{tab:QtQt}
  \end{center}
\end{table}
The top row in Table~\ref{tab:QtQt} shows results without cascade for three different $p_\perp$ correlation assumptions. $Q_\perp$ clearly grows for less local $p_\perp$ conservation, with a relative difference of 25\% between complete $p_\perp$ compensation by neighbours (\textsc{Jetset} default) and uncorrelated $p_\perp$. Adding a cascade almost wipes out the difference, but after a two-jet cut it is restored to a satisfactory degree. 
Also partial $p_\perp$ correlation as assumed in~\cite{jim} gives results after a two-jet cut which differ significantly from the results of the \textsc{Jetset} default assumption.

A comparison of the first and second column illustrates the difference in $p_\perp$ compensation in the cascade and in the fragmentation. The same $p_\perp$ model, tuned with different cascade cut-offs, give significantly different results on $Q_\perp$. 

\section{Example 2: Helix string}\label{sec:screw}
At the end of a perturbative cascade the running coupling becomes relatively large and interference and coherence effects are expected to be
important. In ref~\cite{screw} arguments are presented for a helix-like correlation between
rapidity and azimuth for the gluons at the end of the cascade. 
A modification to the Lund fragmentation scheme reflecting this correlation is presented and the pitch of the helix is given by an unknown parameter $\Delta y/\Delta\phi=\tau$, with expected values around $0.3 -0.5$.

A measure constructed to give a signal for a helix correlation
is presented in ref~\cite{screw}. It is called ``screwiness'' and is defined by
\eqbe S(\omega) \propto \left<\biggl|\sum_{|y_i|<y\srm{cut}}\exp(i(\omega y_i - \phi_i))\biggr|^2\right>, \eqen
where $y_i$ and $\phi_i$ is the rapidity and azimuthal angle for particle $i$, and the sum goes over all particles in a central region specified by the rapidity $y_{\mathrm{cut}}$. If a correlation as in the helix model appears in hadronization, $S(\omega)$ will peak at $\omega\approx 1/ \tau$.

MC simulations of \pair q strings without cascade show a clear signal in $S(\omega)$ for $\tau\sim 0.5$ or larger~\cite{screw}. To check the influence of relatively soft gluons on $S(\omega)$, we have combined the helix fragmentation with the \textsc{Jetset} default treatment of strings with gluons~\cite{gluonkinks}. 

\begin{figure}[t]
  \begin{center}  \hbox{ \vbox{
	\mbox{\psfig{figure=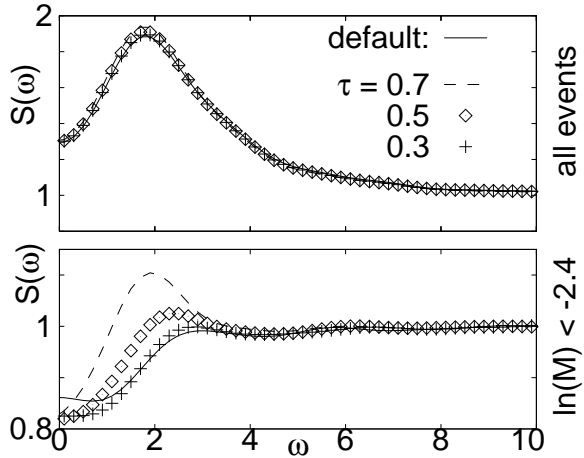,width=0.99\linewidth}}
   }  }
  \end{center}  \caption{\em Screwiness after gluon cascade. Without two-jet cuts (upper plot), long-range correlations in three-jet structures give a signal independent of fragmentation mechanism. After a two-jet event cut (lower plot), this signal is removed, and the $S(\omega)$ measure is sensitive to helix-like fragmentation.}
  \label{f:cutscrew} 
\vspace {5ex}
\end{figure}
The results are shown in  Fig~\ref{f:cutscrew}.
Perturbative activity  introduces long-range azimuthal correlations which give a signal in $S(\omega)$, independent of fragmentation mechanism (upper plot).
After event cuts (lower plot), the observable is more sensitive to the helix model assumptions and gives a clear signal for helix-like string fragmentation if $ \tau\sim 0.5$ or larger.\endcolumn

\section{Summary}~\label{sec:concl}
In many cases the effects of gluon emission overshadow the features of the hadronization phase. To investigate the hadronization mechanism, we have examined how well different strong two-jet cuts suppress the perturbative activity. MC simulations show that no cut can be called optimal. Instead the preferred cut depends on the observable under investigation.

As two examples, we discuss the observables $p_{\perp}$ transfer, sensitive to the locality of $p_{\perp}$ conservation, and ``screwiness'', designed to signal a possible helix correlation in rapidity and azimuth.
After a two-jet cut, we find these observables sensitive to hadronization assumptions. 

The very high statistics now available from experiments at the ${\mrm Z}^0$ resonance implies that it is possible to apply strong two-jet cuts on the data without losing the statistical significance. Our results show that event samples obtained by such cuts can be used to discriminate between different hadronization models and thus be a tool for a more detailed study of the hadronization mechanism.


\begin{thebibliography}{99}
\bibitem{2jet} \bibl{P. Ed\'en, G. Gustafson}{Eur.\ Phys.\ J.}{C8}{1999}{435}
\bibitem{lundstring}  \bibl{B. Andersson {\it et al.}}{Phys.\ Rep.} {97} {1983} {31}
\bibitem{herwig}  \bibl{G. Marchesini {\it et al.}}{Comp.\ Phys.\ Comm.}{67}{1992}{465}
\bibitem{ariadne}   \bibl{L. L\"onnblad}{Comp.\ Phys.\ Comm.}{71} {1992}{15}
\bibitem{jetset}  \bibl{M. Bengtsson, T. Sj\"ostrand}{Comp.\ Phys.\ Comm.}{39}{1986}{347};  \bibl{T. Sj\"ostrand}{Comp.\ Phys.\ Comm.}{82}{1994}{74}
\bibitem{cdm}  \bibl{G. Gustafson}{Phys.\ Lett.\ }{B175}{1986}{453};  \bibl{G. Gustafson, U. Pettersson}{Nucl.\ Phys.\ }{B306}{1988}{746};  \bibl{B. Andersson, G. Gustafson, L. L\"onnblad}{Nucl.\ Phys.\ }{B339}{1990}{393} 
\bibitem{diclus}  \bibl{L. L\"onnblad}{Z. Phys.\ }{C58}{1993}{471}
\bibitem{WolfQt} E.A. de Wolf, private communications.
\bibitem{K2}   E.A. de Wolf, Contrib.\ to {\it Proc.\ XXII Symp.\ on Multiparticle Dynamics} 1992, ed.\ A. Pajares, World Scientific
\bibitem{jim}  \bibl{B. Andersson, G. Gustafson, J. Samuelsson}{Z. Phys.}{C64}{1994}{653}
\bibitem{ucla} \bibl{C.D. Buchanan, S.B. Chun}{Phys.\ Rev.\ Lett.}{59}{1987}{1997}; \bibl{C.D. Buchanan, S.B. Chun}{Phys.\ Rep.}{292}{1998}{239} 
\bibitem{screw}   \bibl{B. Andersson {\it et al.}}{JHEP}{09}{1998}{014}
\bibitem{gluonkinks}   \bibl{T. Sj\"ostrand}{Nucl.\ Phys.}{B248}{1984}{469}

\end{thebibliography}
\end{document}